\documentclass[%
reprint,
superscriptaddress,
 amsmath,amssymb,
prl
floatfix,
]{revtex4-2}

\usepackage{graphicx}
\usepackage{dcolumn}
\usepackage{bm}
\usepackage{amsmath}
\usepackage{comment}
\usepackage{ulem}
\usepackage{xspace}
\usepackage{multirow}
\usepackage{booktabs}

\usepackage[colorlinks,linkcolor=blue,citecolor=blue,anchorcolor=blue]{hyperref}

\def\CeCoInZn{{ CeCo(In$_{1-x}$Zn$_x$)$_5$ }}

\begin{document}

\title{Magnetic Quantum Criticality inside the Superconducting State Revealed by Penetration Depth Scaling with Local $T_{\mathrm c}$}



\author{Yusuke Iguchi}
\email{yiguchi.phys@gmail.com}
\affiliation{Stanford Institute for Materials and Energy Sciences, SLAC National Accelerator Laboratory, 2575 Sand Hill Road, Menlo Park, California 94025, USA}
\affiliation{Geballe Laboratory for Advanced Materials, Stanford University, Stanford, California 94305, USA}

\author{Kaede Inoh}
\affiliation{College of Science, Ibaraki University, Mito, Ibaraki 310-8512, Japan}

\author{Ryosuke Koizumi}
\affiliation{College of Science, Ibaraki University, Mito, Ibaraki 310-8512, Japan}

\author{Makoto Yokoyama}
\affiliation{College of Science, Ibaraki University, Mito, Ibaraki 310-8512, Japan}
\affiliation{Research and Education Center for Atomic Sciences, Ibaraki University, Tokai, Ibaraki 319-1106, Japan}

\date{\today}

\begin{abstract}
We demonstrate a magnetic quantum critical point embedded within the superconducting state of Zn-doped CeCoIn$_5$, revealed by a pronounced peak in the magnetic penetration depth at zero temperature $\lambda(0)$. Using scanning SQUID microscopy, we determine the local superconducting transition temperature $T_{\mathrm c}$ and $\lambda(0)$. By parameterizing $\lambda(0)$ in terms of the local $T_{\mathrm c}$ rather than nominal Zn substitution, we circumvent the ambiguity caused by doping inhomogeneity and enable a more precise extraction of the critical exponent. The extracted exponent exceeds the clean spin-density-wave value, indicating a disorder-modified quantum critical regime. The enhancement of $\lambda(0)$ reflects the suppression of the superfluid stiffness and is consistent with critical scaling. Our approach provides a route to uncover intrinsic quantum critical behavior hidden by inhomogeneity in unconventional superconductors.
\end{abstract}

\maketitle


Quantum critical points (QCPs), continuous phase transitions at zero temperature, are a central organizing principle in correlated electron systems~\cite{SachdevQPT, Gegenwart2008}. In heavy-fermion compounds, antiferromagnetic quantum critical points  (AF-QCPs) are particularly well studied due to the competition between Kondo screening and magnetic order~\cite{Lohneysen2007}. When superconductivity develops in a dome-like region centered around an AF-QCP, unconventional pairing mediated by critical spin fluctuations can emerge~\cite{Mathur1998}. Understanding this interplay remains a central open question.

A key theoretical prediction is that magnetic penetration depth at zero temperature $\lambda(0)$ is enhanced near an AF-QCP by quantum critical fluctuations associated with magnetic order~\cite{Levchenko2013, Chowdhury2013}. While theory predicts a nonanalytic increase of $\lambda(0)$ on the superconducting side of the QCP, whether this leads to an actual peak remains debated~\cite{Chowdhury2013}. Experimentally, a sharp peak in $\lambda(0)$ has been reported in BaFe$_2$(As$_{1-x}$P$_x$)$_2$~\cite{HashimotoScience2012}, and a similar enhancement has been inferred in \CeCoInZn from muon spin relaxation ($\mu$SR)~\cite{HigemotoPNAS2022}.

CeCoIn$_5$ is a prototypical strong-coupling {\it d}-wave superconductor~\cite{Petrovic2001,AnPRL2010}. Zn substitution has been proposed to locally suppresses Kondo screening, thereby inducing AF order and leading to a low-field AF-QCP at extremely low doping levels ($\sim 1\%$)~\cite{HigemotoPNAS2022}. By contrast, the pristine compound exhibits a distinct field-tuned QCP~\cite{Paglione2003} described within an itinerant spin fluctuation (SCR) framework~\cite{SCR,Sakai2011}, suggesting a different microscopic origin for the Zn-induced QCP. The microscopic origin of the Zn-induced AF-QCP remains under discussion: previous work has considered both a Kondo-breakdown scenario and a more itinerant spin-density-wave (SDW)-like scenario, with the latter argued to be more consistent with the absence of clearly divergent mass renormalization in bulk thermodynamic measurements~\cite{HigemotoPNAS2022}. We also note that scanning tunneling microscopy studies on \CeCoInZn films found no clear evidence for impurity-induced AF droplets around individual Zn sites, pointing to a more subtle local effect of Zn substitution~\cite{Haze2018}.

\begin{figure}[!tb]
\includegraphics*[width=8.5cm]{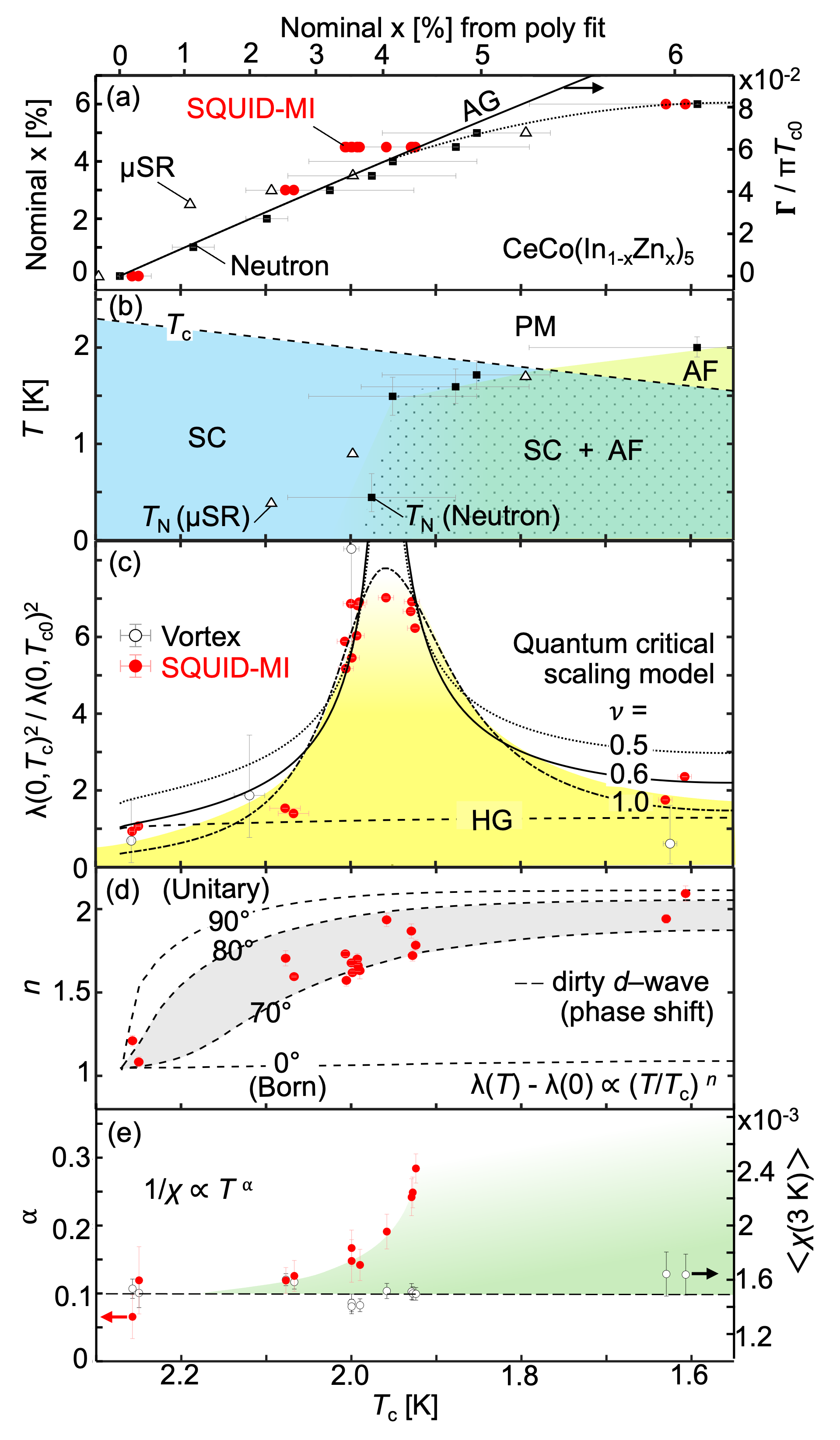}
\caption{\label{fig:Phase_diagram} 
Antiferromagnetic quantum criticality in the dirty {\it d}-wave superconductor \CeCoInZn.
(a) Nominal $x$ is rescaled to $\Gamma/\pi T_{\mathrm c0}$, where $\Gamma$ is the effective impurity scattering strength and $T_{\mathrm c0}$=2.27~K for pristine CeCoIn$_5$. $T_{\mathrm c}$ follows Abrikosov--Gor'kov (AG) theory~\cite{AGtheory} in the superconducting (SC) region and deviates in the coexisting SC and AF region. $T_{\mathrm c}$ data are taken from Refs.~\cite{HigemotoPNAS2022,InohPRB2025}. The top axis indicates the nominal $x$ inferred from the polynomial fit (dashed line), and is shown as a visual guide for panels (b)--(e).
(b) Phase diagram as a function of $T_{\mathrm c}$, showing $T_{\mathrm c}$ and N\'eel temperature $T_{\mathrm N}$ from Refs.~\cite{HigemotoPNAS2022,InohPRB2025}.
(c) $\lambda(0)$ compared with a quantum critical scaling model~\cite{Chowdhury2013} and a strong coupling {\it d}-wave model calculated by Hirschfeld and Goldenfeld (HG)~\cite{HGtheory}. For the quantum-critical scaling model, representative fits obtained with fixed $\nu$ are shown for comparison.
(d) Low-temperature penetration depth exponent $n$ analyzed using a dirty {\it d}-wave model with a finite scattering phase shift~\cite{dirty-d_theta}.
(e) Phenomenological PM susceptibility exponent $\alpha$ and average susceptibility $\chi$ around 3~K. 
}
\end{figure}

However, the extremely low doping levels required to access the Zn-induced AF-QCP regime make spatial inhomogeneity unavoidable. Bulk probes such as $\mu$SR necessarily average over macroscopic volumes, making it difficult to disentangle intrinsic critical behavior from disorder-induced broadening. This is evident in \CeCoInZn, where $\mu$SR measurements on a cluster of crystals~\cite{HigemotoPNAS2022} and neutron scattering on a single crystal~\cite{InohPRB2025} yield substantially different relationships between nominal doping $x$, the superconducting transition temperature $T_{\mathrm c}$, and the N\'eel temperature $T_{\mathrm N}$ [Figs.~\ref{fig:Phase_diagram}(a) and ~\ref{fig:Phase_diagram}(b)]. 

Here we overcome this limitation using scanning superconducting quantum interference device (SQUID) magnetometry and susceptometry to probe superconducting and magnetic properties locally in \CeCoInZn single crystals. This allows simultaneous determination of the local $\lambda(0)$ and local $T_{\mathrm c}$, and we use the latter as an effective tuning parameter [Fig.~\ref{fig:Phase_diagram}]. In contrast to nominal doping, this local $T_{\mathrm c}$ parametrization yields a clearer and more direct mapping of $\lambda(0)$ across the quantum critical regime, avoiding ambiguities associated with doping.

In Fig.~\ref{fig:Phase_diagram}(c), by correlating $\lambda(0)$ with the locally determined $T_{\mathrm c}$, we construct a high-resolution map of the penetration depth across the phase diagram. We observe a pronounced enhancement of $\lambda(0)$ near the AF-QCP, at the local $T_{\mathrm c}$ where $T_{\mathrm N}$ vanishes [Fig.~\ref{fig:Phase_diagram}(b)]. In a scaling analysis with respect to the peak in $\lambda(0)$, the data indicate effective critical behavior that deviates from the clean itinerant SDW expectation, consistent with a disorder-modified regime featuring enhanced local magnetic correlations.

Our results provide a quantitative test of the predicted penetration depth enhancement near an AF-QCP in a heavy-fermion superconductor and establish local probes as a powerful approach for studying quantum criticality in intrinsically inhomogeneous systems.

\begin{figure*}[tb]
\includegraphics*[width=13cm]{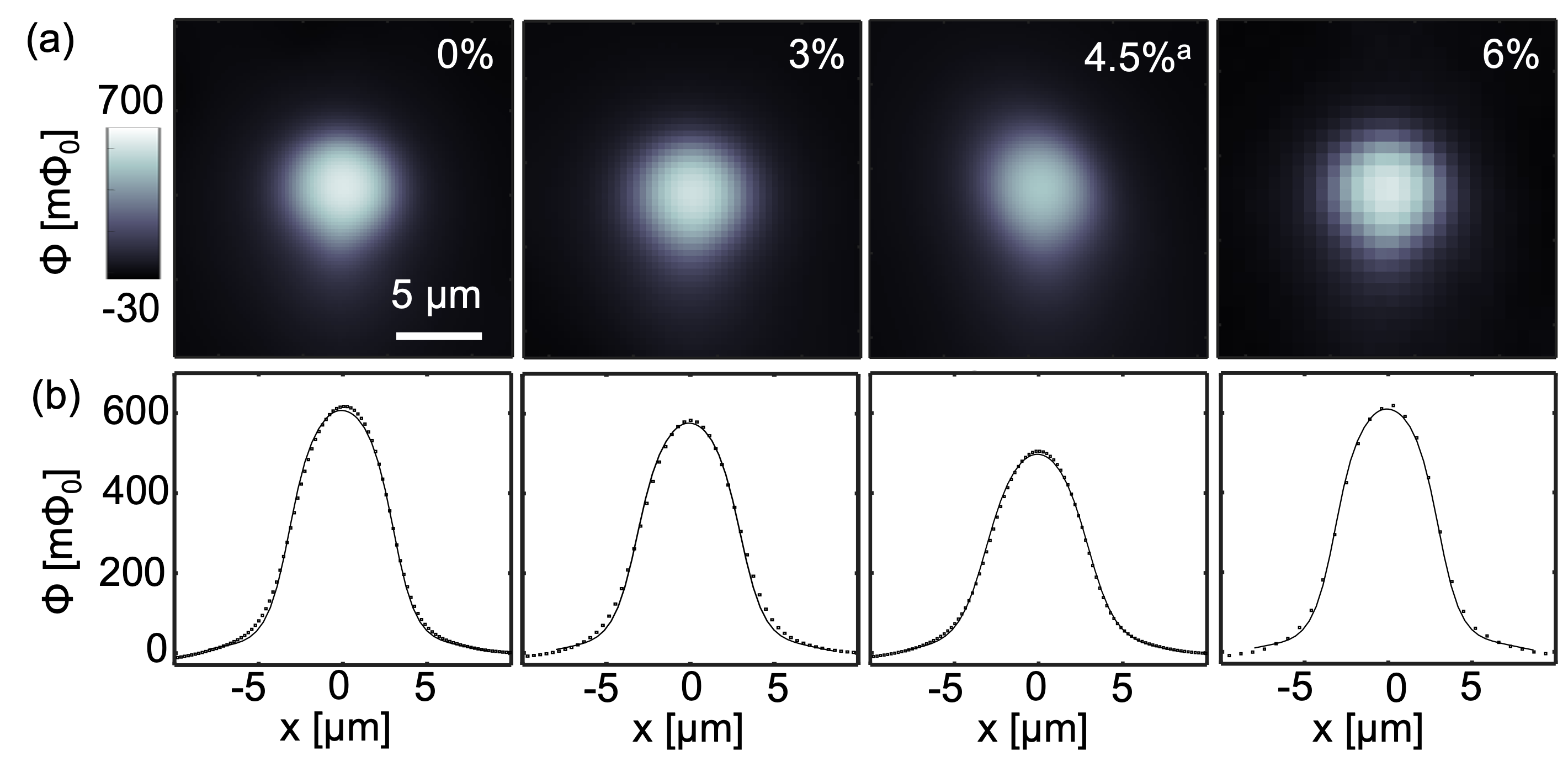}
\caption{\label{fig:vortex} Scanning SQUID magnetometry over an isolated vortex visualizes the enhancement of the penetration depth $\lambda$ near AF-QCP of \CeCoInZn.
(a) Magnetic flux scan over an isolated vortex around 60~mK. (b) The horizontal cross-section of (a) fits the monopole model with the fitting parameter $\lambda$. The smallest peak height at $x=0.045^a$ indicates the enhancement of $\lambda$ near AF-QCP.
}
\end{figure*}

Single crystals of \CeCoInZn (nominal $x \le 0.06$) were grown by the indium-flux method~\cite{Yokoyama2014,Yokoyama2015}. The actual Zn concentration is approximately $0.35x$ based on energy-dispersive x-ray spectroscopy, but we use the nominal concentration $x$ throughout. The $ab$-plane surfaces were hand polished to $\sim$50~nm final grit. The data reported here were obtained from five samples with nominal Zn concentrations $x=0, 0.03, 0.045^a, 0.045^b$, and 0.06.

Local magnetic measurements were performed using a scanning SQUID microscope in a dilution refrigerator down to $\sim 60$~mK. The sensor consists of a gradiometric pickup loop/field coil with inner radii of $\sim 3.0~\mu$m and $\sim 1.5~\mu$m, respectively~\cite{Kirtleyrsi2016}. Magnetometry images the local dc flux $\Phi$ from isolated vortices in units of the flux quantum $\Phi_0=h/2e$. Susceptometry applies an ac current $I^{\mathrm{ac}} \sim 1$~mA at $\sim 1$~kHz to the field coil and detects the induced ac flux $\Phi^{\mathrm{ac}}$. At contact, the corresponding maximum excitation field at the sample surface is of order $0.1$~mT. We define the local susceptibility as the mutual inductance (MI) $M = \Phi^{\mathrm{ac}}/I^{\mathrm{ac}}$ in units of $\Phi_0$/A.


To track the penetration depth across the AF-QCP in \CeCoInZn, we first imaged isolated vortices at $T\sim 60$~mK in samples with nominal $x=0$, 3\%, $4.5\%^a$, and 6\% [Fig.~\ref{fig:vortex}(a)]. The vortex profiles show a clear doping dependence, with a strongly suppressed peak height at $x=0.045$. We fit the data using the monopole approximation for the vortex stray field~\cite{Kirtley1999}, $B_z(x,y) = \Phi_0(z_s+\lambda)/(x^2+y^2+(z_s+\lambda)^2)^{3/2}$,
taking $\lambda$ as the only free parameter and fixing the scan height to $z_s=800\pm100$~nm. We simulated the total flux through the pickup loop area from this field profile. As shown in Fig.~\ref{fig:vortex}(b), the fits describe the vortex cross sections well, yielding $\lambda=170$, 280, 590, and $160\pm50$~nm for nominal $x=0$, 0.03, $0.045^a$, and 0.06, respectively. The extracted $\lambda$ shows a pronounced maximum near $x=0.045$ [Fig.~\ref{fig:Phase_diagram}(c)], close to the AF-QCP identified by neutron scattering~\cite{InohPRB2025}. Since these data were acquired at 60 mK ($\sim3\%$ of $T_c$), they provide an excellent approximation to $\lambda(0)$ within our experimental resolution.

For a more systematic determination of the local penetration depth and $T_{\mathrm c}$, we then performed scanning SQUID susceptometry on the same regions. Compared with vortex imaging, susceptometry enables efficient local measurements of the Meissner response below $T_{\mathrm c}$, the paramagnetic (PM) susceptibility above $T_{\mathrm c}$, and the local $T_{\mathrm c}$ itself, all with micron-scale resolution.

\begin{figure}[tb]
\begin{center}
\includegraphics*[width=8cm]{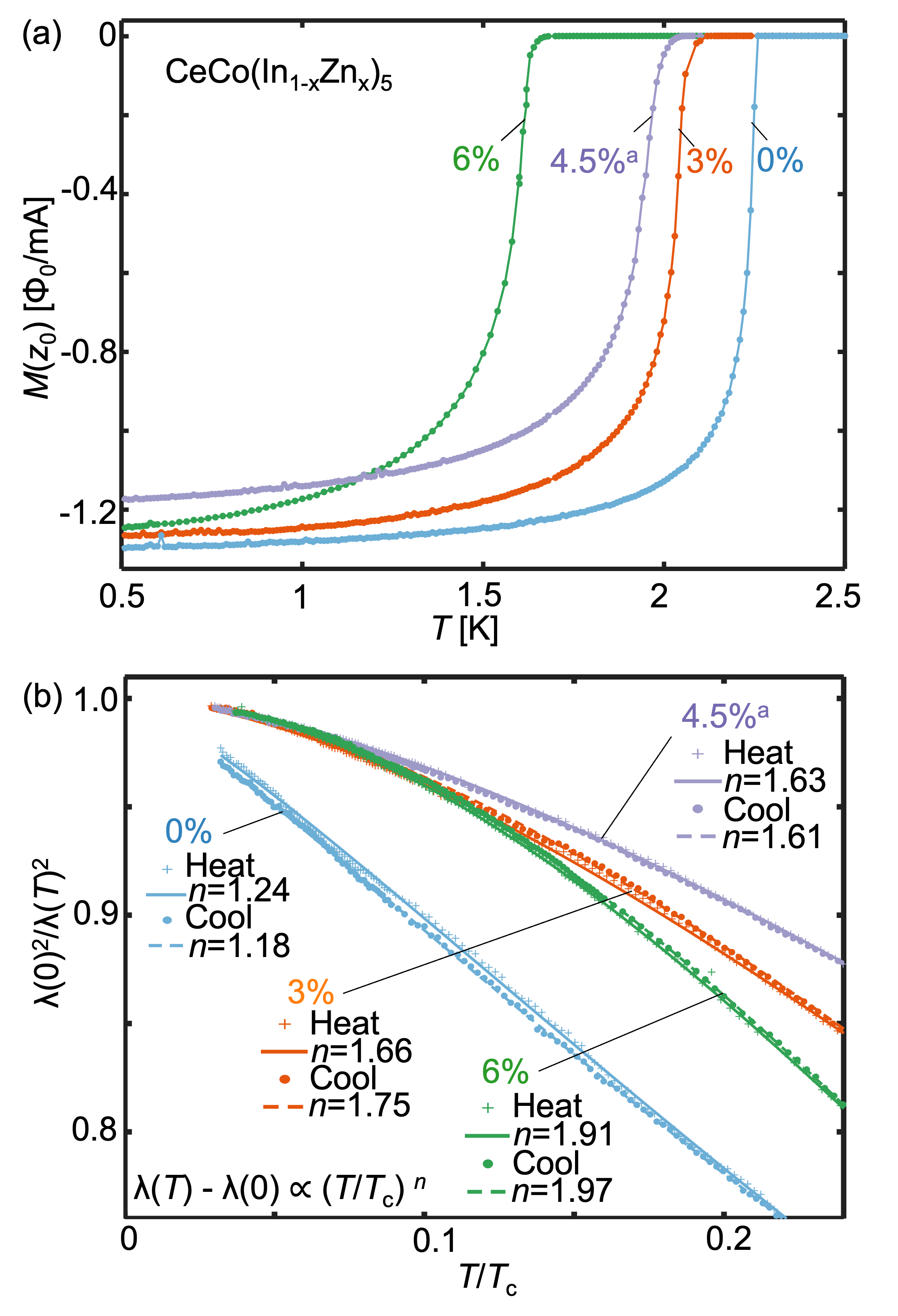}
\caption{\label{fig:Sup-T} Scanning SQUID Susceptometry at low temperatures in \CeCoInZn reveals an enhancement of $\lambda(0)$ toward AF-QCP and an increase in the low-temperature exponent $n$ with Zn doping. (a) The temperature dependence of $M(z=z_0)$ at several locations of $x=0B, 0.03B, 0.045^aD, 0.06B$ (see Fig.~S6 for additional locations; local $T_{\mathrm c}$ values for each location are listed in supplemental Table~I). The SQUID sensor just touches the sample surface at $z_0$. (b) Superfluid density $\lambda(0)^2/\lambda(T)^2$ at low temperatures. The data are analyzed using the power-law form $\lambda(T) - \lambda(0) = (T/T_{\mathrm c})^n$. The solid and dashed lines are fits to the heating and cooling data, respectively. The uncertainty in $n$ is dominated by the hysteresis between heating and cooling.
}
\end{center}
\end{figure}

We analyze the susceptometry signal using established models for diamagnetic screening and PM response~\cite{Kirtley2012}, extracting the local $\lambda$ and magnetic volume susceptibility $\chi$ (see Supplemental Material~\cite{supple}). We define the local $T_{\mathrm c}$ by $M(z=z_0)=-50~\Phi_0/\mathrm{A}$, where $z_0$ is the SQUID height at contact. Spatial maps of $T_{\mathrm c}$ [Fig.~S2] reveal measurable microscopic inhomogeneity in the doped samples. At fixed SQUID height $z_0$, the temperature dependence of $M(z_0)$ [Fig.~\ref{fig:Sup-T}(a)] captures both the local onset of superconductivity and the evolution of $\lambda$; in particular, the reduced low-temperature magnitude near $x=0.045$ indicates weaker diamagnetic screening and hence a larger penetration depth near the AF-QCP.

Above $T_{\mathrm c}$, the temperature dependence of the PM susceptibility $\chi(T)$ evolves systematically with Zn concentration, with the magnitude of its increase upon cooling varying accordingly [Fig.~S4]. We analyze $\chi(T)$ between $T_{\mathrm c}$ and 3~K using a phenomenological form, $1/\chi \propto T^{\alpha}$, which yields a non-Curie-like exponent $\alpha<1$ [Fig.~\ref{fig:Phase_diagram}(e)]. The exponent $\alpha$ increases rapidly with doping, whereas the average susceptibility $\chi$ at 2.8--3~K remains nearly unchanged.

Despite this evolution, no sharp anomaly is resolved in $M(z_0,T)$ near the expected $T_{\mathrm N}$ for either $x=0.06$ or $x=0.045$ (see Supplemental Material~\cite{supple}). For $x=0.06$, where neutron scattering reported AF order with the $c$ axis as the easy axis~\cite{InohPRB2025}, the broad temperature dependence of our out-of-plane response is qualitatively similar to the $c$-axis susceptibility reported for Zn-doped CeCoIn$_5$ in bulk measurements~\cite{Yokoyama2015}. Together with the precursor-like enhancement in the specific heat above $T_{\mathrm N}$ reported in the same work~\cite{Yokoyama2015}, this suggests that AF correlations develop already above $T_{\mathrm N}$ and produce a broad precursor-like suppression of the susceptibility rather than a sharp peak. Such behavior is also consistent with the nanoscale magnetic inhomogeneity inferred from nuclear magnetic resonance measurements~\cite{Sakai2021}. For $x=0.045$, the absence of a distinct feature may similarly reflect either weak coupling between AF fluctuations and Cooper pairs or a broadened precursor response associated with incipient AF order.

To further investigate the low-energy superconducting properties, we extracted the normalized superfluid density $\lambda(0)^2/\lambda(T)^2$ and fit the low-temperature variation using the phenomenological form $\lambda(T) - \lambda(0) = (T/T_{\mathrm c})^n$, where $n$ reflects both the gap structure~\cite{Gross1986} and nonmagnetic impurity scattering in a {\it d}-wave superconductor (dirty {\it d}-wave model)~\cite{AGtheory,HGtheory} [Fig.~\ref{fig:Sup-T}(b) and Fig.~S3]. In the pristine compound, we obtain $n = 1.05--1.24$ depending on location, close to the linear-$T$ behavior expected for a clean {\it d}-wave state. Previous reports for CeCoIn$_5$ span $n = 1.25$~\cite{Kim2015,Truncik2013} to $n = 1.5$~\cite{Chia2003,Ozcan2003,HashimotoPNAS2013}, the latter having been discussed in terms of quantum-critical renormalization of nodal quasiparticles~\cite{HashimotoPNAS2013}. The broad spread of reported values, however, indicates that the low-temperature exponent is not universal even in the pristine CeCoIn$_5$. By contrast, our data show a systematic increase of $n$ with Zn substitution [Fig.~\ref{fig:Phase_diagram}(d) and Fig.~\ref{fig:Sup-T}(b)], which is robust and broadly consistent with increasing impurity scattering in a dirty {\it d}-wave state.

The extracted $\lambda(0)$ provides an independent estimate of the penetration depth, complementing the vortex-based values in Fig.~\ref{fig:vortex} [Fig.~\ref{fig:Phase_diagram}(c)]. For pristine CeCoIn$_5$, we obtain $\lambda(0)=200\pm$100~nm, consistent with previous microwave and two-coil measurements (196--281~nm)~\cite{Ormeno2002,Ozcan2003,Truncik2013}, though smaller than $\mu$SR estimates of 550--611~nm~\cite{Higemoto2002,HigemotoPNAS2022}. 
Figure~\ref{fig:Phase_diagram}(b) shows the phase diagram versus $T_{\mathrm c}$, which serves as an effective tuning parameter by reducing the effects of doping inhomogeneity compared with nominal substitution. In this representation, the AF-QCP identified by neutron scattering lies near $T_{\mathrm c} \sim 1.95$~K, where both vortex fitting and susceptometry reveal a pronounced enhancement of $\lambda(0)$ [Fig.~\ref{fig:Phase_diagram}(c)].

To estimate the effective impurity scattering strength $\Gamma$, we use the Abrikosov--Gor'kov expression for dirty {\it d}-wave superconductors~\cite{AGtheory}, $\log\!\left(T_{\mathrm c}/T_{\mathrm c0}\right)=\Psi\!\left(1/2\right)-\Psi\!\left(1/2+\Gamma /2\pi T_{\mathrm c0}\right)$, where $\Psi$ is the digamma function and and $T_{\mathrm c0}$ is $T_{\mathrm c}$ of pristine CeCoIn$_5$. Fitting the suppression of $T_{\mathrm c}$ in the superconducting-only regime ($T_{\mathrm c}>1.9$~K) [Fig.~\ref{fig:Phase_diagram}(a)] yields a critical scattering strength corresponding to an effective Zn concentration $x \sim 0.072$, i.e., $\sim35\%$ of the nominal content, implying a critical nominal concentration of $x \sim 0.21$. This is comparable to the actual critical dopant concentration in the related nonmagnetic impurity system CeCo(In$_{1-x}$Sn$_x$)$_5$ ($x \sim 0.093$), which does not exhibit magnetic order~\cite{BauerPRB2006}. With this impurity scale, the doping evolution of $n$ is broadly captured by a dirty {\it d}-wave model with the scattering phase shift $\theta = 70$--$80^\circ$
~\cite{dirty-d_theta}, i.e., near the unitary limit [Fig.~\ref{fig:Phase_diagram}(d)], consistent with previous studies on electron-irradiated YBa$_2$Cu$_3$O$_{7-x}$, which also found $\theta \sim70^\circ$ \cite{dirty-d_theta}. Near the AF-QCP, however, $n$ should be regarded as an effective exponent shaped by both disorder and critical magnetic fluctuations. In this sense, the exponent $n$ is a less direct measure of the underlying quantum-critical singularity than $\lambda(0)$.

In contrast, the peak in $\lambda(0)^2$ is incompatible with the monotonic increase expected from the strong-coupling dirty {\it d}-wave model with unitary impurity scattering~\cite{HGtheory},
$\lambda(T=0,\Gamma)^2/\lambda(0,0)^2
= 1 + (0.79)^2 \Gamma \pi/3 + 0.79\sqrt{\pi\Gamma/3} = h(\Gamma)$,
and its magnitude far exceeds what can be accounted for by impurity scattering alone [see the line labeled HG in Fig.~\ref{fig:Phase_diagram}(c)]. To quantify the enhancement, we fit Fig.~\ref{fig:Phase_diagram}(c) to the scaling form predicted near a magnetic QCP~\cite{Chowdhury2013}:
$ \lambda(0,\Gamma)^2 = 1/(C + A|\Gamma - \Gamma_c|^{3\nu - 1})=q(\Gamma,\Gamma_c,A,C,\nu)$, where the magnetic correlation length scales as $\xi\sim|\Gamma-\Gamma_c|^{-\nu}$. To incorporate both quantum-critical scaling and impurity scattering, we fit the measured $\lambda(0,T_\mathrm{c})^2$ using $q(\Gamma,\Gamma_c,A,C,\nu)h(\Gamma)$, with $C$, $A$ and $\nu$ as fitting parameters and $\Gamma_c$ fixed at nominal $x=$4\%, corresponding to the observed center of the $\lambda(0,T_\mathrm{c})^2$ peak (see Supplemental Material~\cite{supple}). 
Within this experimentally motivated peak-centered parametrization, the data disfavor the clean Gaussian value $\nu=0.5$ expected for an itinerant SDW QCP within the Hertz--Millis framework~\cite{Lohneysen2007}, and instead favor larger effective exponents. 
This is naturally compatible with a disorder-modified critical regime, since quenched disorder is expected to be a relevant perturbation to the clean Hertz fixed point in $d=2$ and $3$. Any resulting conventional dirty critical point must then satisfy the Harris criterion, $\nu \ge 2/d$~\cite{Lohneysen2007}.

Real-space inhomogeneity is intrinsic to doping-tuned QCPs and cannot be eliminated even with a local probe. However, the use of locally determined $T_\mathrm{c}$ substantially reduces the artificial broadening associated with bulk averaging over nominal doping and allows the critical enhancement of $\lambda(0)^2$ to be resolved more faithfully. The remaining width of the peak reflects intrinsic real-space inhomogeneity of the sample, implying that the measured $\lambda(0)^2$ still provides a spatially averaged response and thus a lower bound on the intrinsic divergence. Within this framework, the data are most naturally interpreted in terms of a disorder-broadened critical regime with enhanced local magnetic correlations beyond the simplest Hertz--Millis description. 

The phenomenological susceptibility exponent $\alpha$ shows a different evolution. Above $T_\mathrm{c}$, $1/\chi \propto T^\alpha$ yields $\alpha<1$, and $\alpha$ increases strongly with Zn content while the average magnitude of $\chi$ remains nearly unchanged [Fig.~\ref{fig:Phase_diagram}(e)]. This behavior is not expected for a simple clean quantum-critical scenario with universal exponents, nor is it consistent with a trivial increase of a Curie term from a growing number of fully localized moments. Rather, the systematic growth of $\alpha$ is more naturally interpreted as a disorder-sensitive magnetic response amplified near the AF instability, for example through impurity-enhanced AF correlations and/or Griffiths-like magnetic fluctuations~\cite{Stewart2001}. Similar disorder-sensitive, nonuniversal magnetic exponents have been discussed in disordered Kondo-lattice systems such as CePd$_{1-x}$Rh$_x$~\cite{CePdRhGriffiths} and in disordered quantum critical systems such as Ni$_{1-x}$V$_x$~\cite{UbaidKassis2010}. In this sense, the Zn evolution of $\alpha$ is consistent with the disorder-broadened critical regime inferred independently from the local $\lambda(0)$ analysis, although the present data do not by themselves distinguish among these scenarios or establish a Griffiths phase. Thus, $\alpha$ behaves as a more disorder-sensitive effective exponent than $\lambda(0)$, which provides a more direct measure of the critical suppression of the superfluid stiffness.

The larger value of $\alpha$ reported for CeCoIn$_5$ in bulk measurements under 0.1~T may indicate some field sensitivity of the anomalous susceptibility, although the absolute magnitude of $\chi$ around 3~K is compatible with our results~\cite{KimPRB2001}. However, this field scale is much smaller than that relevant to the field-tuned quantum critical regime ($\sim$5~T) in pristine CeCoIn$_5$~\cite{Paglione2003}, and therefore does not by itself imply proximity to a field-driven QCP.


We have used scanning SQUID microscopy to probe the superconducting properties of \CeCoInZn at the microscale and determined both the local penetration depth $\lambda(0)$ and the local superconducting transition temperature $T_{\mathrm c}$. By using the local $T_{\mathrm c}$ rather than nominal doping as the tuning parameter, we reduce ambiguity from spatial inhomogeneity and thereby resolve the enhancement of $\lambda(0)$ near the AF-QCP more quantitatively. By correlating local measurements of $T_{\mathrm c}$, $\lambda(0)$, $\lambda(T) \propto T^n$, and $1/\chi(T) \propto T^\alpha$, we show that the sharp peak in $\lambda(0)^2$ cannot be explained by impurity scattering alone, but is instead consistent with quantum-critical enhancement. Within an experimentally motivated peak-centered scaling analysis, the data disfavor the clean Gaussian value $\nu=0.5$ expected from conventional itinerant SDW theory and instead favor larger effective exponents, indicating that disorder and local magnetic correlations are important ingredients of the critical regime. Rather than eliminating intrinsic inhomogeneity, the role of the local probe is to reduce additional broadening from macroscopic averaging and thereby expose the disorder-modified quantum critical regime realized in the actual doped material. Our results establish the combination of local penetration-depth measurements and local $T_{\mathrm c}$ mapping as a powerful approach for probing quantum-critical behavior in intrinsically inhomogeneous superconductors.

\begin{acknowledgments}
We thank Kathryn A. Moler and Youichi Yanase for fruitful discussion.
This work was primarily supported by the DOE “Quantum Sensing and Quantum Materials” Energy Frontier Research Center under Grant No. DE-SC0021238. Sample synthesis at Ibaraki was supported by Japan Society for the Promotion of Science KAKENHI Grants No. 23K03315.
\end{acknowledgments}


\newpage
\clearpage

\setcounter{figure}{0}
\setcounter{equation}{0}
\renewcommand{\thefigure}{S\arabic{figure}}
\renewcommand{\theequation}{S\arabic{equation}}

\onecolumngrid
\appendix
	
\begin{center}
	\Large
	{Supplemental Material for \\\lq\lq Magnetic Quantum Criticality inside the Superconducting State Revealed by Penetration Depth Scaling with Local $T_{\mathrm c}$ \rq\rq} \\by Iguchi $et$ $al.$
\end{center}
	
\subsection{Scan Height Dependence of Scanning SQUID Susceptometry}
\begin{figure*}[tb]
\begin{center}
\includegraphics*[width=16.5cm]{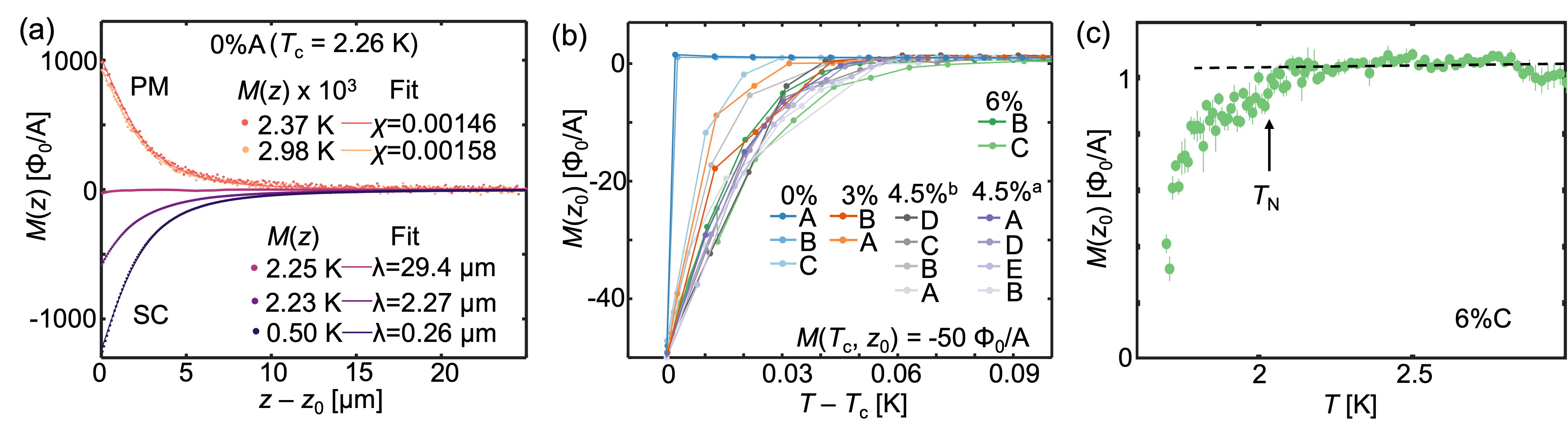}
\caption{\label{fig:SSS} Scanning SQUID susceptometry microscopically estimates the local $T_{\mathrm c}$ and the penetration depth enhanced toward AF-QCP of CeCo(In$_{1-x}$Zn$_{x}$)$_5$.
(a) The SQUID height dependence of the mutual-inductance $M(z)$ at the location A of $x=0$ sample at each temperature. The curves fit the magnetic susceptibility $\chi$ above $T_{\mathrm c}$ and the penetration depth $\lambda$ below $T_{\mathrm c}$. (b) $M(z_0)$ plotted as a function of $T-T_{\mathrm c}$ for different Zn concentrations and locations. (c) Paramagnetic susceptibility $M(z_0)$ for $x=0.06$ starts decreasing above $T_{\mathrm c}$, indicating Ne\'{e}l temperature $T_N$. 
}
\end{center}
\end{figure*}

Figure~\ref{fig:SSS}(a) shows the mutual inductance $M(z)$ as a function of SQUID height for various temperatures. The data are well fit by a model incorporating either the diamagnetic screening (parameterized by $\lambda$) or a paramagnetic susceptibility $\chi$, depending on whether $T$ is below or above $T_{\mathrm c}$, respectively~\cite{Kirtley2012}. The local $T_{\mathrm c}$ is defined as $M(z_0) = -50~\Phi_0$/A [Fig.~\ref{fig:SSS}(b)]. The onset of the diamagnetic response is sharp in the parent compound, whereas it becomes more gradual in the doped samples. This broadened evolution of the susceptometry signal suggests the presence of superconducting inhomogeneity on a length scale smaller than the micron-scale spatial resolution of scanning SQUID susceptometry.

In addition, we observe a systematic increase in paramagnetic susceptibility above $T_{\mathrm c}$ with increasing Zn concentration. In the x=0.06 sample, where $T_{\mathrm N}>T_{\mathrm c}$, $M(z_0)$ exhibits neither a clear peak at $T_{\mathrm N}$ nor a pronounced decrease below it [Fig.~\ref{fig:SSS}(c)]. Since neutron scattering has reported antiferromagnetic order with the $c$ axis as the easy axis~\cite{InohPRB2025}, a more distinct anomaly would be expected in our out-of-plane susceptometry geometry for a homogeneous long-range ordered state. Its absence suggests that the magnetic state is instead short-ranged or strongly inhomogeneous, more in line with the NMR results~\cite{Sakai2021}. In the $x=0.045$ sample, where $T_{\mathrm N}<T_{\mathrm c}$ is expected, no anomaly is detected either, implying that any coupling between antiferromagnetic fluctuations and Cooper pairs is too weak to produce a resolvable susceptometry feature.

\subsection{Fitting Analysis of the penetration depth enhancement at AF-QCP}

To examine how the extracted critical exponent depends on the fitting procedure, we analyze the $\lambda(0)^2$ data using the combined form introduced in the main text,
\begin{equation}
\lambda(0,\Gamma)^2 = q(\Gamma,\Gamma_c,A,C,\nu)\,h(\Gamma),
\end{equation}
where
\begin{equation}
q(\Gamma,\Gamma_c,A,C,\nu)
=
\frac{1}{C + A|\Gamma-\Gamma_c|^{3\nu-1}}
\end{equation}
describes the quantum-critical enhancement~\cite{Chowdhury2013}, and
\begin{equation}
h(\Gamma)
=
1 + (0.79)^2 \Gamma \pi/3 + 0.79\sqrt{\pi\Gamma/3}
\end{equation}
describes the strong-coupling dirty {\it d}-wave contribution in the unitary limit~\cite{HGtheory}.

In this parametrization, $A$ controls the overall strength of the critical enhancement, while $C$ sets the scale that limits the apparent divergence at the QCP and hence controls the peak amplitude. For a given $\nu$, $C$ varies monotonically together with $A$, so the fit landscape is most transparently visualized in the $(\nu,A)$ plane. Therefore, Fig.~\ref{fig:Sim-lam}(a) shows the residual sum of squares (RSS) in the $(\nu,A)$ plane with $C$ fixed to its best-fit value.

The global minimum is obtained at $\nu=0.922$, $A=639$, and $C=0.15$. Figure~\ref{fig:Sim-lam}(a) plots the RSS normalized by its minimum value, together with contours corresponding to 5\%, 10\%, 20\%, and 50\% increases above the minimum. Figure~\ref{fig:Sim-lam}(b) compares the experimental data with the global best fit and with fits obtained by fixing $\nu=1.0$, 0.8, 0.7, 0.6, and 0.5 and optimizing the remaining parameters.

In performing this analysis, we exclude one point just above the peak center (nominal $x=0.04$), where the sample shows the strongest microscopic inhomogeneity and the local estimate of $\lambda$ is least reliable. We also fix $\Gamma_c$ to nominal $x=0.04$, corresponding to the observed center of the $\lambda(0)^2$ peak. This choice is phenomenological but not arbitrary: shifting $\Gamma_c$ away from this value leads to visibly poorer fits. Moreover, nearly symmetric peak-like enhancements of $\lambda$ have been reported in other quantum-critical superconductors~\cite{HashimotoScience2012}, even though their microscopic origin remains under debate. In this sense, the present peak-centered parametrization is experimentally motivated, while the resulting preference for $\nu>0.5$ should still be understood within this fitting framework.

A further complication is that, in the coexistence regime of superconductivity and AF order, the Abrikosov-Gor'kov description of $T_{\mathrm c}$ suppression is no longer adequate, because AF order competes with superconductivity. Therefore, to display the fitted curves in Fig.~\ref{fig:Sim-lam}(b) on the nominal-$x$ axis, we convert the measured local $T_{\mathrm c}$ to nominal $x$ using a polynomial interpolation of the experimental $x$-$T_{\mathrm c}$ relation rather than the Abrikosov-Gor'kov fit [Fig.~1(a)]. This introduces a slight asymmetry in the displayed curves, which reflects the experimental mapping between $T_{\mathrm c}$ and nominal $x$, rather than an intrinsic asymmetry of the underlying scaling form.

The main conclusion of this analysis is that the clean Gaussian value $\nu=0.5$ cannot reproduce the sharp rise of $\lambda(0,\Gamma)^2$ near the AF-QCP within the present peak-centered fitting procedure, whereas larger effective exponents, already from $\nu \sim 0.6$, provide substantially better agreement. The global best fit occurs near $\nu \sim 0.9$, but the precise optimum should not be overinterpreted, because the fit quality depends in part on how the residuals are distributed between the low-$x$ side and the peak region. In particular, the rounded peak structure observed experimentally can arise both from intrinsic scaling with larger effective $\nu$ and from residual real-space inhomogeneity. Therefore, this analysis should be taken as evidence that the data disfavor $\nu=0.5$ within a simple clean SDW-like description, rather than as a precise or model-independent determination of $\nu$.

\begin{table}[htbp]
\centering
\caption{Local $T_c$ of CeCo(In$_{1-x}$Zn$_x$)$_5$ estimated from scanning SQUID susceptometry measurements in Fig.~\ref{fig:SSS}. While the fitting procedure yields sub-mK numerical precision, the effective uncertainty is limited by the temperature step ($\sim$10 mK).}
\label{tab:localTc}
\begin{tabular}{c||ccc|cc|ccccccc|cccc|cc}
\cline{1-19}
\toprule
\multirow{1}{*}{$x$}
  & \multicolumn{3}{c|}{0}
  & \multicolumn{2}{c|}{0.03}
  & \multicolumn{7}{c|}{0.045$^a$}
  & \multicolumn{4}{c|}{0.045$^b$} 
  & \multicolumn{2}{c}{0.06} \\
Loc  & A & B & C
  & B & A 
  & G & H & A & D & F & E & B
  & D & C & B & A & B & C \\ 
\midrule
$T_c$(K) 
  & 2.26 & 2.26 & 2.25
  & 2.08 & 2.07
  & 2.01 & 2.01 & 2.00 & 2.00 & 1.99 & 1.99 & 1.99
  & 1.96 & 1.93 & 1.93 & 1.92
  & 1.63 & 1.61 \\
\bottomrule
\cline{1-19}
\end{tabular}
\end{table}

\clearpage

\begin{figure}[htb]
\includegraphics*[width=15.5cm]{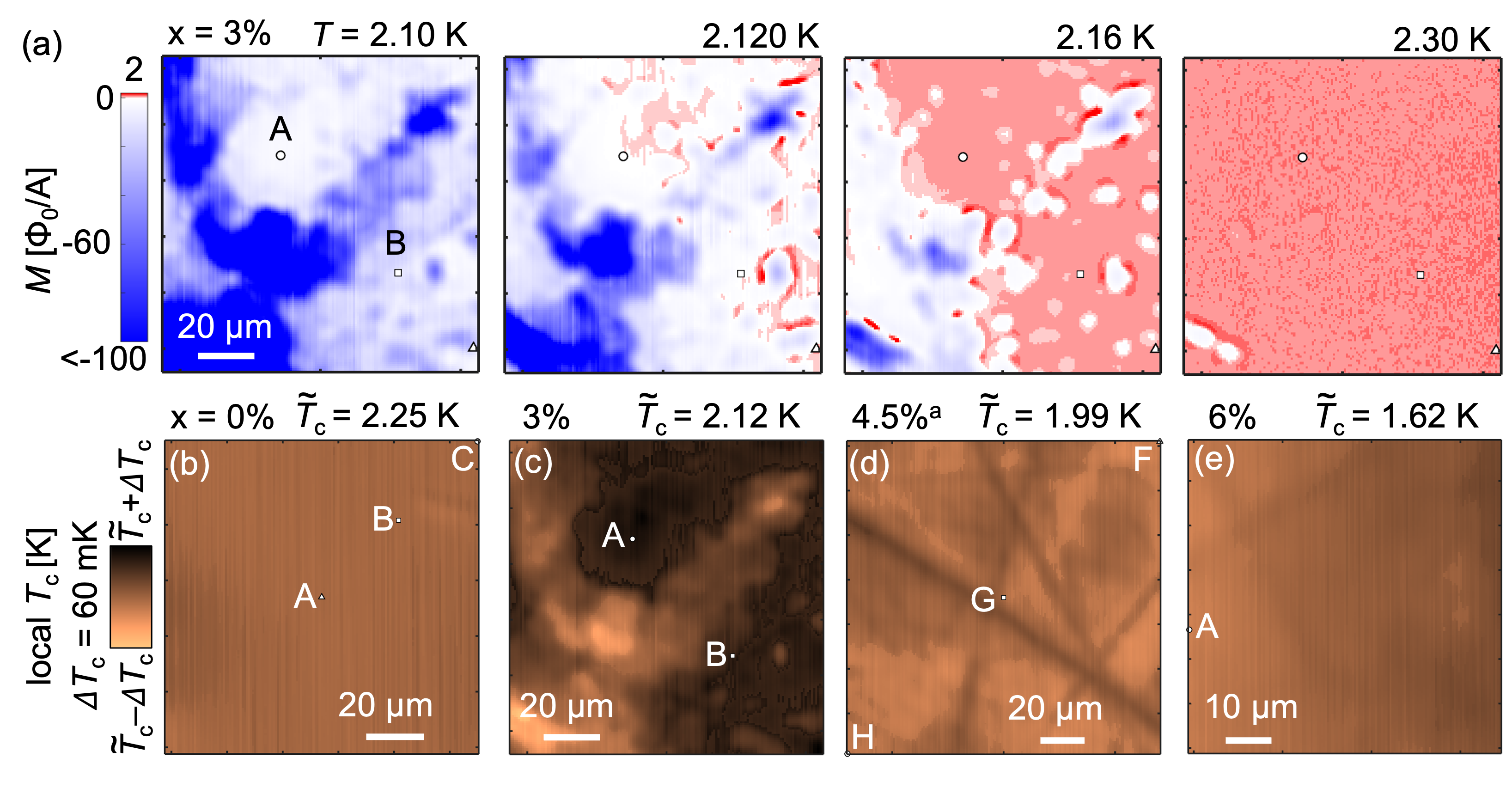}
\caption{\label{fig:SSSscan_Tdep} (a) Susceptibility scan of 3\% sample. (b-e) Spatial mapping of the local superconducting transition temperature $T_{\mathrm c}$, visualizing microscopic inhomogeneity within the scanned area. The local $T_{\mathrm c}$ is defined by $M(T_{\mathrm c}) = -50~\Phi_0/\mathrm{A}$. The variation of $T_{\mathrm c}$ within each region is 20, 120, 50, and 40 mK for panels (a–d), respectively. A common color scale is used for all panels.
}
\end{figure}

\begin{figure}[htb]
\includegraphics*[width=16.5cm]{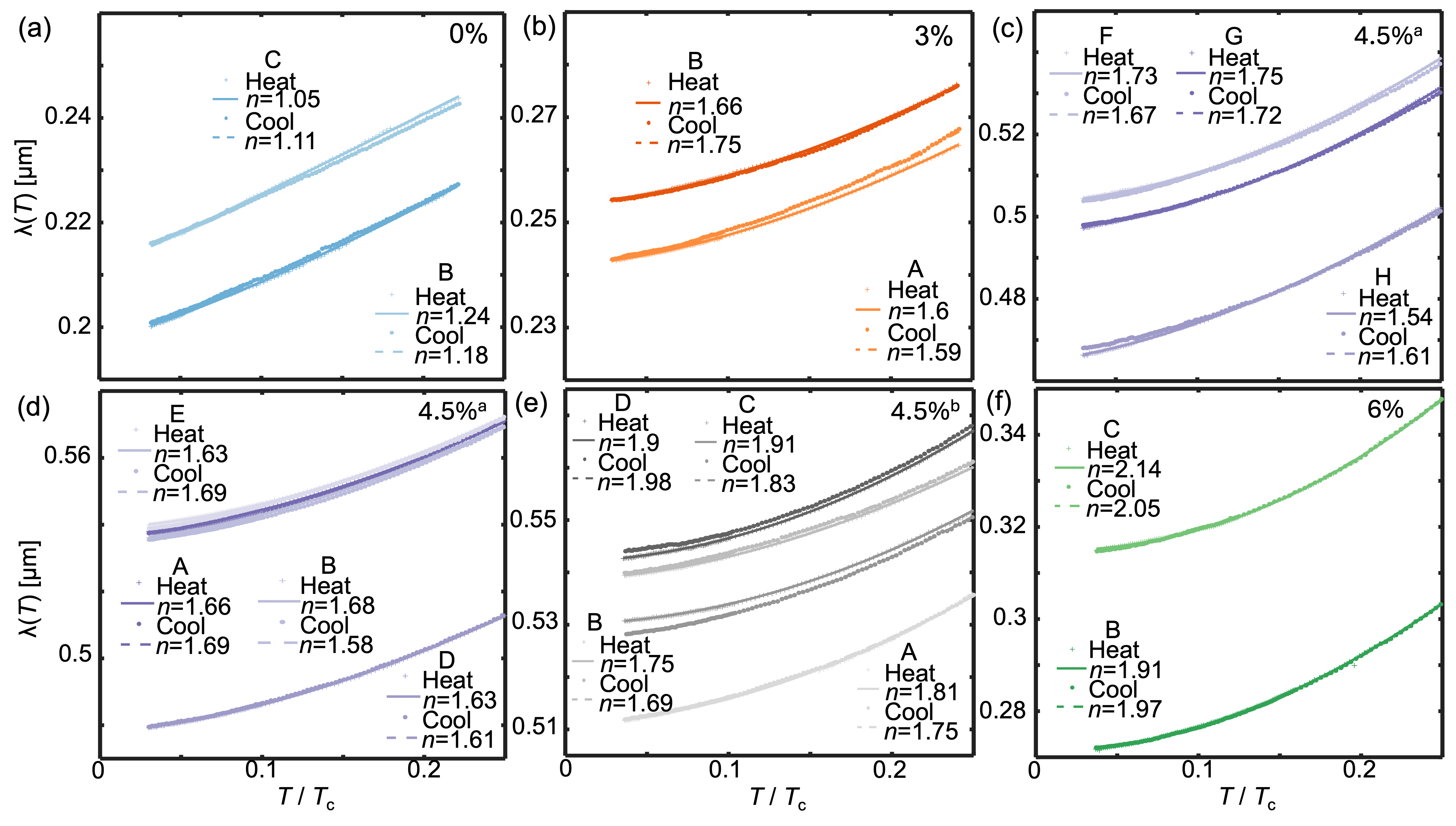}
\caption{\label{fig:pene-T} Temperature dependence of Penetration depth of CeCo(In$_{1-x}$Zn$_{x}$)$_5$ obtained from the time averaged susceptibility $<M(z_0)>_t$ in heating and cooling. 
}
\end{figure}

\begin{figure}[htb]
\includegraphics*[width=16.5cm]{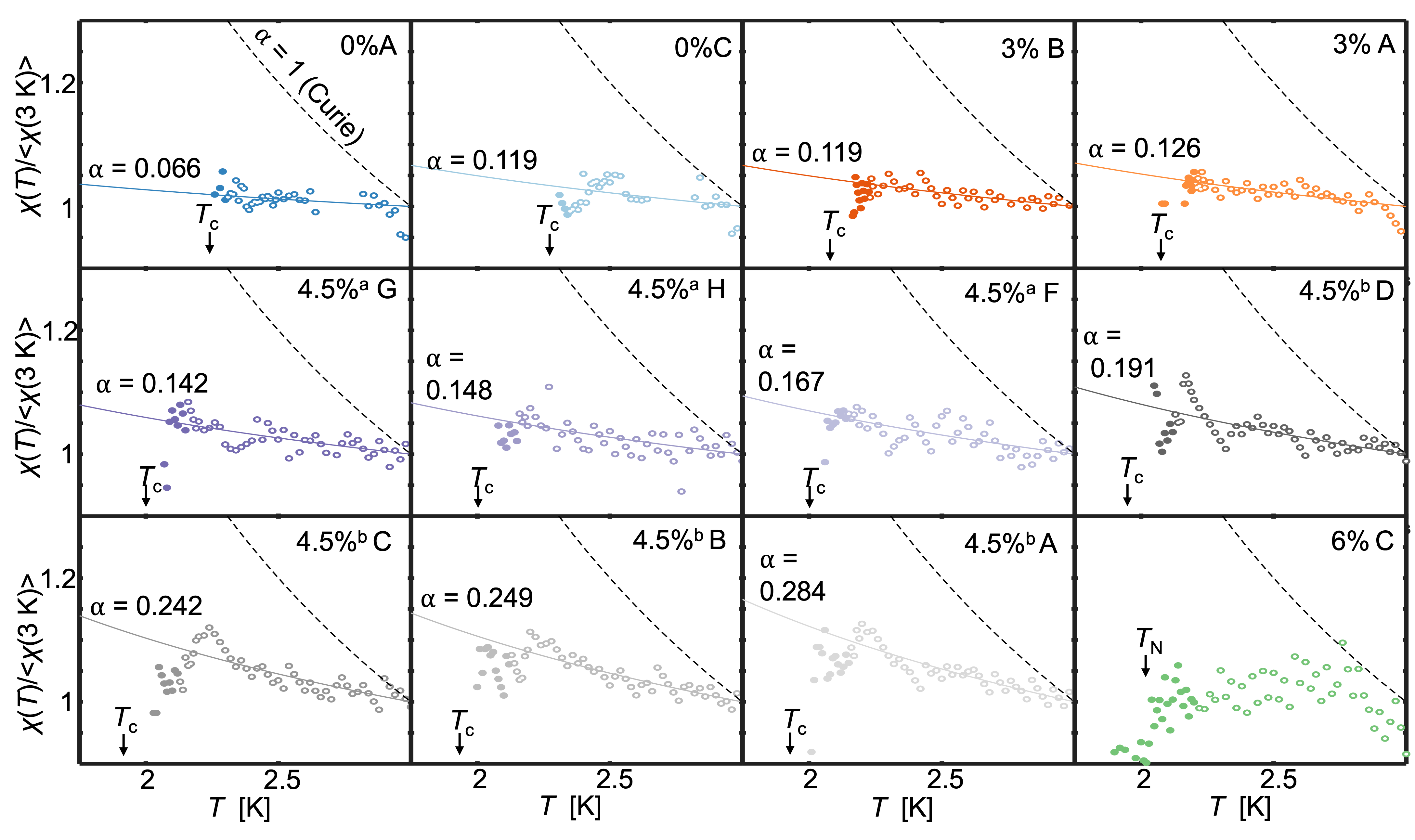}
\caption{\label{fig:PM-T_sep} Temperature dependence of the paramagnetic susceptibility $\chi$, obtained by fitting $M(z)$ (such as Fig. 2(a)), fits the power law in CeCo(In$_{1-x}$Zn$_{x}$)$_5$.
}
\end{figure}

\begin{figure}[htb]
\includegraphics*[width=15cm]{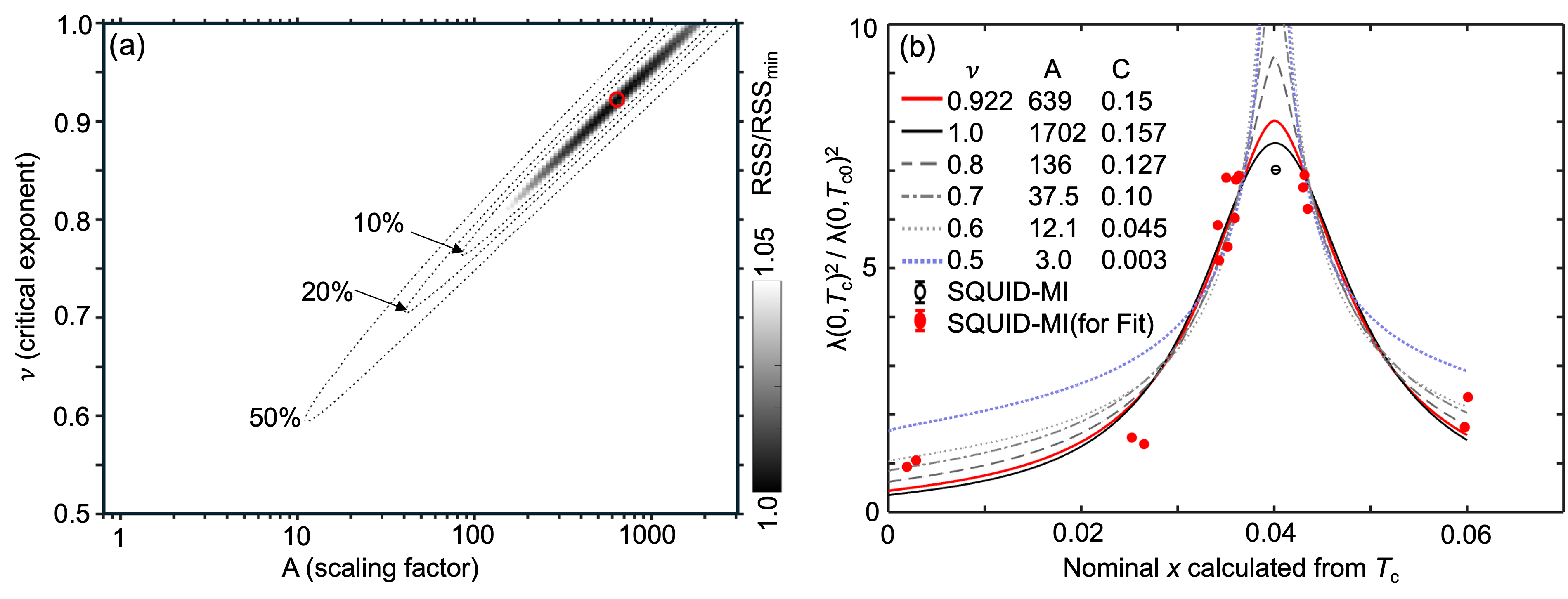}
\caption{\label{fig:Sim-lam} 
Fitting analysis of $\lambda(0)^2$.
(a) Residual sum of squares (RSS), normalized by its global minimum, plotted in the $(\nu,A)$ plane for the quantum-critical scaling fit. The parameter $C$, which controls the peak amplitude by limiting the apparent divergence at the QCP, is fixed to its best-fit value because it varies monotonically together with $A$ over the relevant range. Dashed contours indicate 5\%, 10\%, 20\%, and 50\% increases above the minimum. The red circle marks the global minimum at $\nu=0.922$, $A=639$, and $C=0.15$.
(b) Experimental data compared with the global best fit (red solid line) and with fits obtained by fixing $\nu=1.0$, 0.8, 0.7, 0.6, and 0.5 and optimizing the remaining parameters. The vertical axis is the same normalized quantity \(\lambda(0,T_{\mathrm c})^2/\lambda(0,T_{\mathrm c0})^2\) used in Fig.~1(c) of the main text. For visual convenience, the horizontal axis is plotted as nominal $x$ inferred from the measured local $T_{\mathrm c}$; this representation is equivalent to the $\Gamma$-based fitting used in the text. The point at nominal $x=0.04$ just above the peak center is excluded from the fit because it lies in the region with the strongest microscopic inhomogeneity and the least reliable local estimate of $\lambda$.
}
\end{figure}

\begin{figure}[tb]
\begin{center}
\includegraphics*[width=8cm]{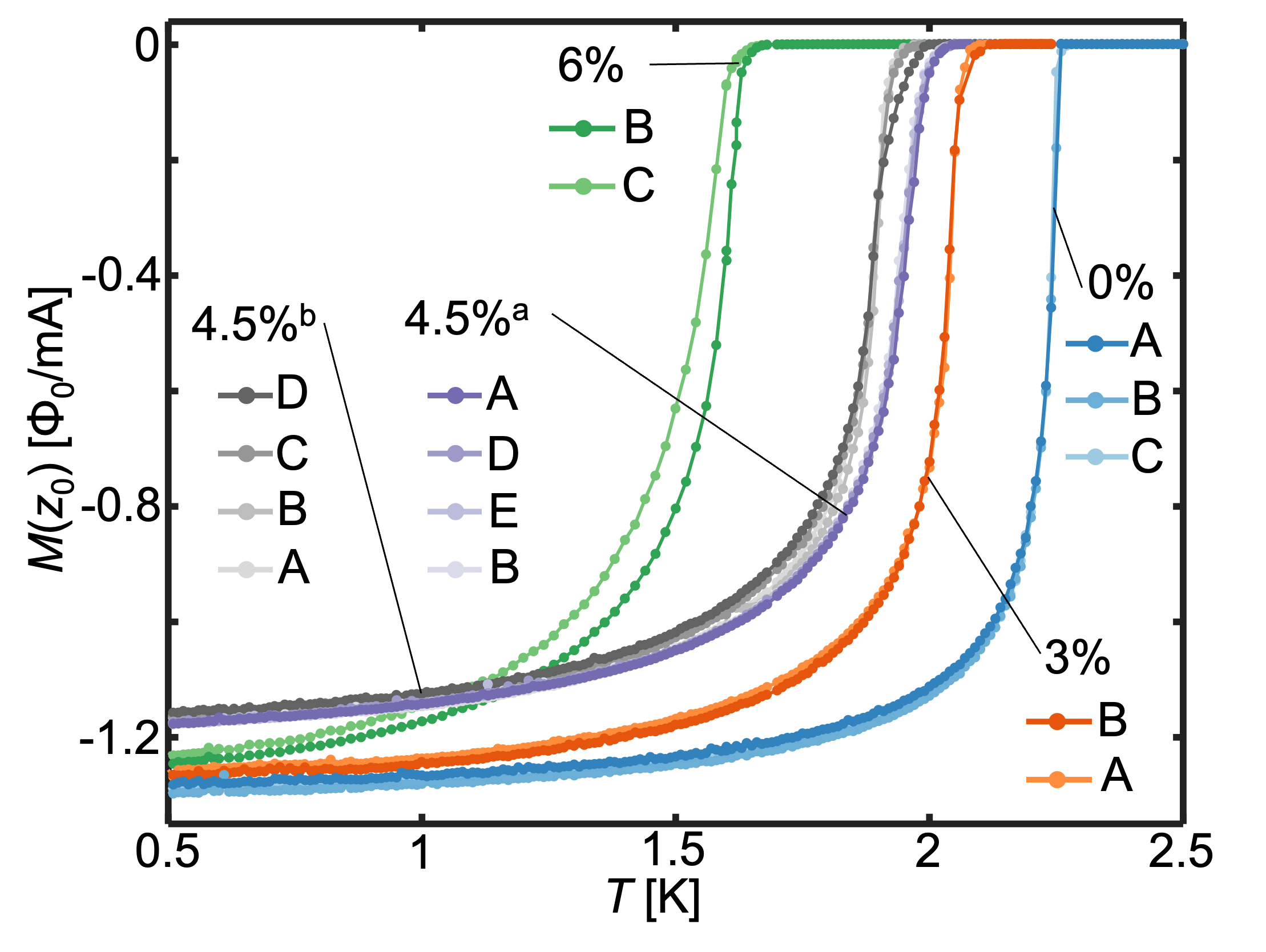}
\caption{\label{fig:sus-Tall} All data of the temperature dependence of $M(z=z_0)$.
}
\end{center}
\end{figure}

\end{document}